\documentclass[useAMS,usenatbib]{mn2e}
\usepackage{graphicx,graphics,aas_macros,amsmath}
\usepackage{hyperref}

\title[OAO 1657--415]{Variations in the pulsation and spectral characteristics of OAO 1657--415.}
\author[P. Pradhan, C. Maitra, B.Paul, N.Islam, B.C. Paul]{Pragati Pradhan$^{1,2}$ \thanks{E-mail:pragati2707@gmail.com;}
 Chandreyee Maitra$^{3,4}$ Biswajit Paul$^{3}$ Nazma Islam$^{3,4}$ B.C. Paul$^{2}$ \\
$^{1}$ St. Joseph's College, Singamari, Darjeeling-734104, West Bengal, India\\
$^{2}$ North Bengal University, Raja Rammohanpur,  District Darjeeling-734013, West Bengal, India \\
$^{3}$ Raman Research Institute, Sadashivnagar, Bangalore-560080, India\\
$^{4}$ Joint Astronomy Programme, Indian Institute of Science, Bangalore-560012, India\\ }
\begin{document}
\date{}
%\pagerange{\pageref{firstpage}--\pageref{lastpage}} \pubyear{2011}
\maketitle
\label{firstpage}
\begin{abstract}
We present broad-band pulsation and spectral characteristics of the accreting
X-ray pulsar OAO 1657--415 with a 2.2 d long \emph{Suzaku} observation carried out
covering its orbital phase range $\sim$ 0.12--0.34, with respect to the mid-eclipse. During the last third of the
observation, the X-ray count rate in both the X-ray Imaging Spectrometer (XIS) and the HXD-PIN instruments
increased by a factor of more than 10. During this observation,
the hardness ratio also changed by a factor of more than 5, uncorrelated with
the intensity variations. In two segments of the observation, lasting for $\sim$
30--50 ks, the hardness ratio is very high. In these segments, the spectrum
shows a large absorption column density and correspondingly large equivalent
widths of the iron fluorescence lines. We found no conclusive evidence for the presence of a cyclotron line in 
the broad-band 
X-ray spectrum with \emph{Suzaku}.
The pulse profile, especially in the XIS energy band shows evolution with time but not so with energy.
We discuss the nature of the intensity variations, and variations of the absorption column density and
emission lines during the duration of the observation as would be expected 
due to a clumpy stellar wind of the
supergiant companion star.  These results indicate that OAO 1657--415 has
characteristics intermediate to the normal supergiant systems and
the systems that show fast X-ray transient phenomena.

\end{abstract} 
\section{Introduction}

\label{sect:intro}
OAO 1657--415 is an accreting binary X-ray pulsar with a pulse period of
$\sim$ 38 \rm{s}  \citep{WP79}  discovered with the \emph{Copernicus}
satellite \citep{P78}. The companion star is an Ofpe/WN9 type supergiant \citep{MA09} which is characterized by slow winds, high mass-loss rates
and exposed CNO-cycle products. 
This binary system has an orbital period of $\sim$10.5 \rm{days} \citep{C93}
with an orbital decay, $\dot{P}_{orb}$ $\sim$ $-9.74\times10^{-8}$ \citep{J12}. 
The moderate value of its spin and orbital period gives it a unique place in
the Corbet diagram \citep{C86} intermediary to the two classes of sources
which transfer mass via stellar wind and Roche lobe overflow \citep{C93}. 
Using \emph{ASCA} observations, a dust scattered halo was found which was used to
estimate the distance of the source as 7.1 $\pm$ 1.3 \rm{kpc} \citep{A06}, consistent with a distance of 6.4 $\pm$ 1.5 {kpc} estimated
earlier based on the study of the pulsar's infrared counterpart \citep{C02}.\\
OAO 1657--415 has shown spin-up/down periods and torque reversals in the past \citep{WP79,C02} 
a phenomenon common to accreting neutron stars \citep{BI97}
which could not be explained by wind accretion \citep{B97}. 
Relation of the pulse frequency evolution with X-ray luminosity is not well
understood \citep{B00} for this pulsar.\\
The light curve of the X-ray source shows a complete eclipse lasting for about $\sim$
$\frac{1}{5}$ of the orbital period. With a very large number of short
exposures with the \emph{INTEGRAL}-IBIS, it was found that outside the eclipse, the
light curve shows another dip in the phase range of 0.55 \citep{BS08}. 
Overall, even outside the eclipse and the dip, the X-ray intensity varies by a factor
of several \citep{BS08}. However, the \emph{INTEGRAL} observations did not establish whether these
observed fluctuations are due to intensity variation from orbit to orbit or
due to large intensity variations within an orbit of the X-ray binary. \\
The X-ray spectrum of OAO 1657--415 is similar to the class of high magnetic
field neutron stars, with a $K{\alpha}$ fluorescence iron
emission line at $\sim$ 6.4 \rm{keV} \cite[]{C02,A06}
along with $K{\beta}$ iron line emission at 7.1 \rm{keV}. 
Being at a low Galactic latitude, or due to large amount of circumstellar material, the X-ray spectrum of OAO 1657--415 is highly absorbed.
Possible existence of a Cyclotron Resonance Scattering Feature feature at $\sim$ 36 \rm{keV}
was seen  with limited statistical significance in the broad-band spectrum obtained with \emph{Beppo}-SAX indicating magnetic
field strength $3.2 (1+z)\times10^{12}$ \rm{G}, where \emph{z} is the gravitational
redshift  \cite[]{O99,BS08}. \\
Here, we report results from an analysis of the broad-band pulsation and
spectral characteristics of the pulsar OAO 1657--415 using a long observation
carried out with the \emph{Suzaku} observatory. The pulse profile and the spectral
parameters of the source are characterized over the duration of the observation.
The time resolved measurement of the absorption column and the iron
fluorescence line intensity are useful to investigate interaction of the
pulsar X-rays with the stellar wind of the companion thereby providing
some wind diagnostics like wind clumpiness. We make a comparison of the intensity and spectral
variability of this source with same of the supergiant fast x-ray
transients (SFXTs). \\
\section{Observation and Data Analysis}
\label{sect:data analysis}
\emph{Suzaku} \citep{M07} is a broad-band X-ray observatory which covers
the energy range of 0.2--600 \rmfamily{keV}. It has two main instruments:
(i) the X-ray Imaging Spectrometer (XIS; \citep{K07} covering 0.2--12
\rmfamily{keV} range and
(ii) the Hard X-ray Detector (HXD) having PIN diodes \citep{T07} covering 
the energy range of 10-70 \rmfamily{keV} and GSO crystal scintillators
detectors covering 70-600 \rmfamily{keV}. 
The XIS consists of four CCD detectors of which three are front illuminated 
 and one is back illuminated. Three out of the four XIS units (XIS 0,1 and 3) are
operational since  2006. \newline
OAO 1657--415 was observed with \emph{Suzaku} during 2011-09-26
(OBSID `406011010'). The observation was carried out at the `XIS nominal'
pointing position and lasted for $\sim 200$ \rm{ks}. The XISs were operated
in `standard' data mode in the `Window 1/4' option which gave a time
resolution of 2 \rm{s}. The MJD of observation is listed in Table \ref{mjd}

\begin{table} 
\scriptsize
\caption{MJD-OBS and useful exposure for each XIS and PIN}
\label{table}
\centering
\begin{tabular}{|c | c | c |}
\hline
Instrument  & MJD-OBS & Useful exposure \\
\hline
XISs  & 55830--55832 & $\sim$ 84.7 \rm{ks} \\
PIN  & 55830--55832 & $\sim$ 75 \rm{ks} \\
\hline
\end{tabular}
\label{mjd}
\end{table}
\noindent
For the XIS and HXD data, we used the filtered cleaned event files which are obtained using the pre-determined screening criteria as given in 
\emph{Suzaku} ABC guide. 
The XIS event files were checked for photon pile-up\footnote{http://www-utheal.phys.s.u-
tokyo.ac.jp/$\sim$yuasa/wiki/index.php\\/How\_to\_check\_pile\_up\_of
\_Suzaku\_XIS\_data}
and were not piled up with the peak count rate per one CCD being $\sim$ 6 \rm{count arcmin$^{-2}$ exposure} when compared with the peak count rate per one
CCD for Crab being $\sim$ 36 \rm{count arcmin$^{-2}$ exposure}. During the last 50 \rm{ks} of the observation, the
light curve shows a large increase in luminosity. 
We also checked for a possible pile-up for this region and did not find any. The 
peak count rate per one CCD for this region being $\sim$ 12 \rm{count arcmin$^{-2}$ exposure}. \\
XIS light curves and spectra were extracted from the XIS data by
choosing circular regions of 3 \rm{arcmin} radius around the source centroid.
Background light curves and spectra were extracted by selecting regions of
the same size away from the source. 
The XIS spectra were extracted with 2048 channels.
The average XIS and PIN count rates were 
2.5 \rm{count s$^{-1}$} and 2.1 \rm{count s$^{-1}$} respectively.
For HXD--PIN background, simulated `tuned' non-X-ray background event files corresponding to the month and year of the respective observations 
were used to estimate the non X-ray
background\footnote{\url{http://heasarc.nasa.gov/docs/suzaku/analysis/pinbgd.html}}
\citep{F09}. 
Response files for the XIS was created using CALDB version `20130916' and for HXD-PIN spectrum, response files corresponding to the epoch 
of observation were obtained from the \emph{Suzaku} guest observatory
facility\footnote{\url{http://heasarc.nasa.gov/docs/heasarc/caldb/suzaku/}}. 
\subsection{Timing Analysis}
\label{sect:timing analysis}
\begin{figure*}
\begin{center}
\includegraphics[width=9cm,height=12cm,angle=-90]{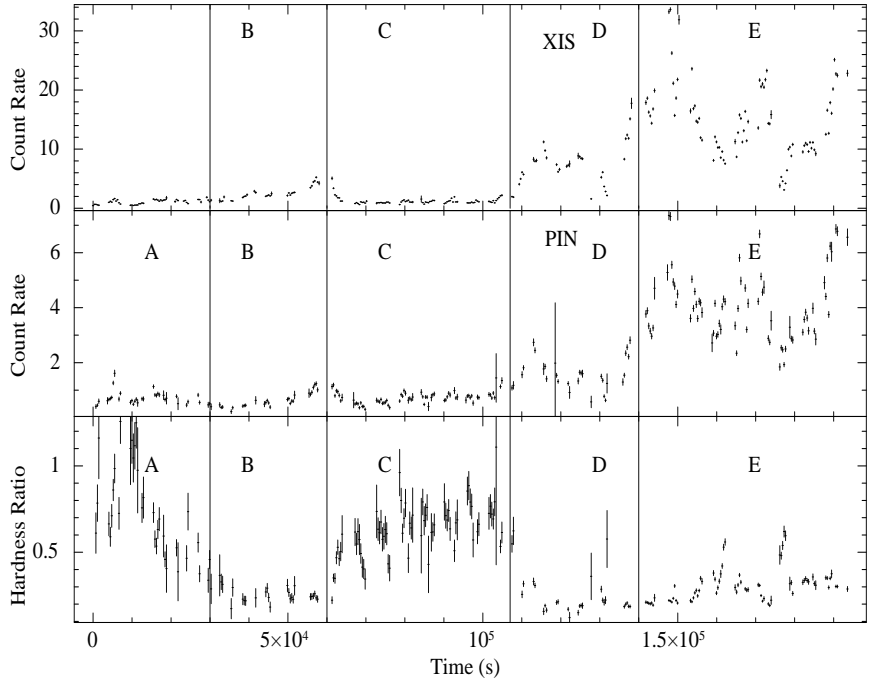}
\end{center}
\caption{\rmfamily{Background-subtracted light curves of OAO 1657--415 with a binning of 10 times the pulsar period (369 \rm{s}). 
The upper and middle panels represent the XIS and PIN data respectively. The lower panel represents the hardness ratio. The
zero in time corresponds to MJD 55830.4 
in this figure and the Figs \ref{npex_he_param_cyclabs} and \ref{eqw12_nh_npex_he}.}}
\label{lcurve} 
\end{figure*}

\begin{figure*}
\begin{center}
\includegraphics[width=9cm,height=12cm,angle=-90]{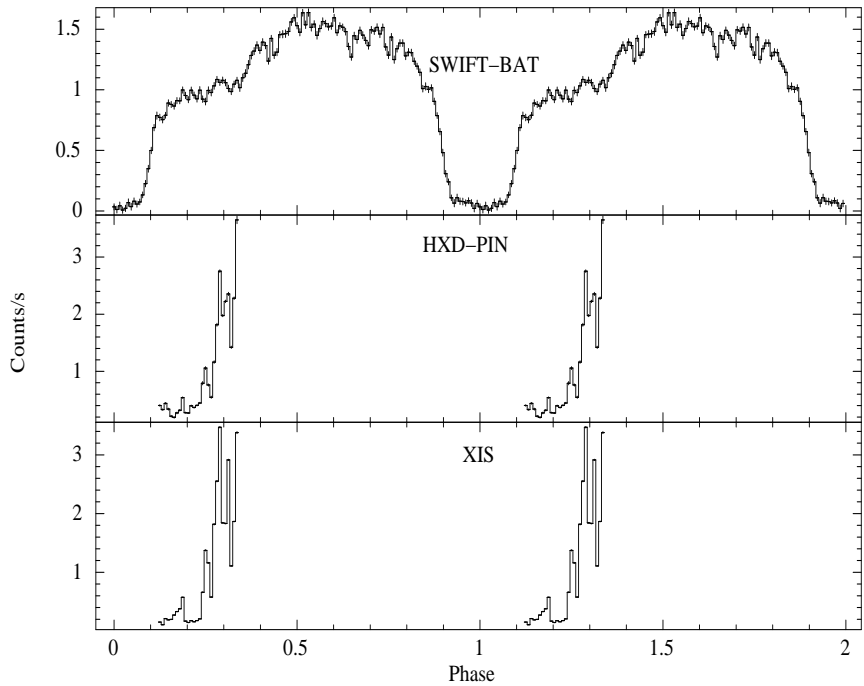}
\end{center}
\caption{{\rmfamily
Light curve of \emph{Swift}-BAT folded with the orbital period of 
OAO 1657--415 and the HXD-PIN and XIS light curves obtained are shown here. 
The phase zero corresponds to the mid-eclipse time MJD 55776.9135.}}
\label{swift-suzaku} 
\end{figure*}

Timing analysis was performed on the XIS and PIN light curves after applying
barycentric corrections to the event data files using the \texttt{FTOOLS} task
`aebarycen' and dead time corrections were done using \texttt{FTOOLS} task `hxddtcor'. Light curves were extracted from the XIS data with the minimum available time resolution of 2 \rm{s}. 
The average exposure time was $\sim$ 85 ks for XIS.
For the PIN data, light curves with a resolution of  1 \rmfamily{s} were extracted, the exposure time being $\sim$ 75 ks. 
We summed the background-subtracted XIS 0, 1 and 3 light curves and obtained a single background-corrected light curve for XISs. 
The PIN light curves were background subtracted by generating a background light curve using the simulated background files. 
For all the timing analysis for XISs and PIN which we discuss throughout the paper, we use these two background subtracted light curves.
A plot of the light curve with binning of 10 times the spin period of the pulsar is shown in 
Fig. \ref{lcurve}. The light curve shows very low count rate during the first 110 \rmfamily{ks}.
In the rest of the observation the count rate increased by a factor of several and there is a 
large variation in count rate.
Though the count rate in this first 110 \rmfamily{ks} is small, it is
still highly variable by a factor of a few.
In Fig. \ref{lcurve}, the upper and middle panel represents the XIS and PIN data respectively. 
The lower panel represents the hardness ratio between the PIN and XIS. 
Based on the hardness ratio, the light curve is divided into five segments (A,B,C,D,E).
\\
\\ \noindent
To investigate the orbital phase of the \emph{Suzaku} observation, we folded the
long-term light curve of OAO 1657--415 obtained with \emph{Swift-BAT}\footnote{\url{http://swift.gsfc.nasa.gov/results/transients/BAT_current.html\#anchor-EXO1657-419}}
at the orbital period of $\sim$ 10.447 \rm{d} \citep{J12} which is shown in Fig.  \ref{swift-suzaku} along with the
\emph{Suzaku}-XIS and PIN light curves. 
 It clearly shows that during the \emph{Suzaku} observation, the source was not in
eclipse.

\subsection{Time- and energy-resolved pulse profiles}
\begin{figure*}
\includegraphics[scale=0.5,angle=-90]{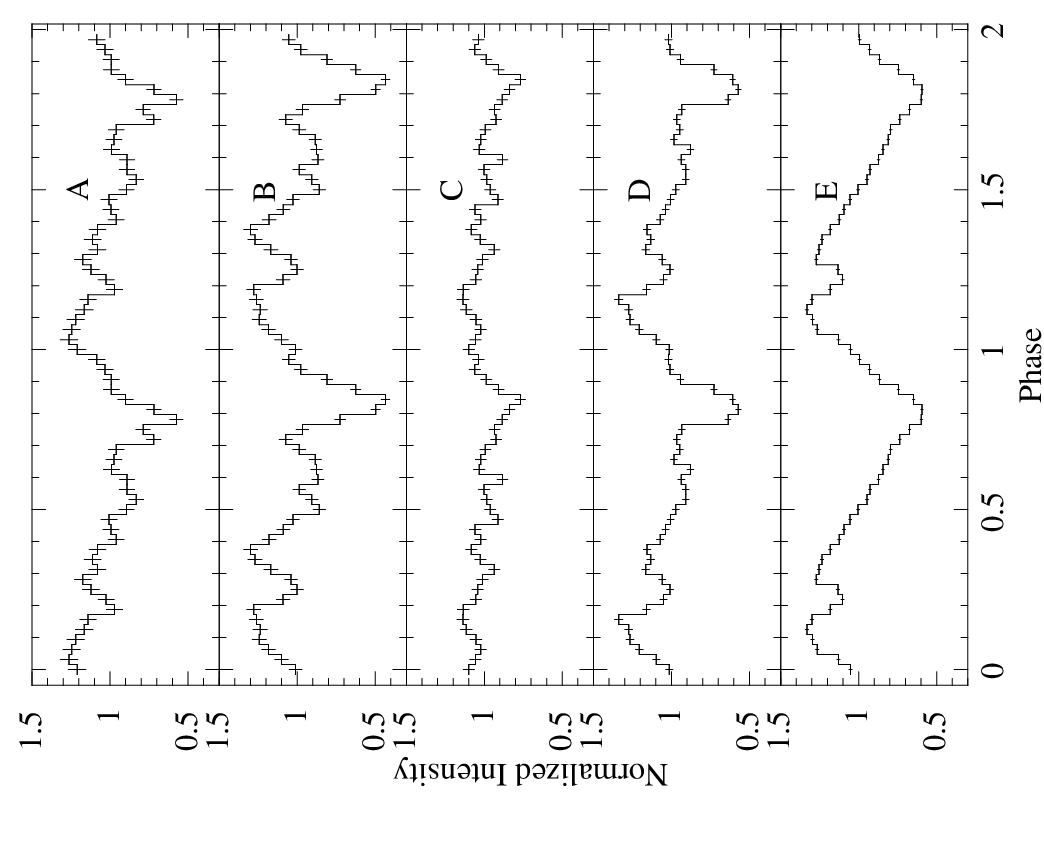}
\includegraphics[scale=0.5,angle=-90]{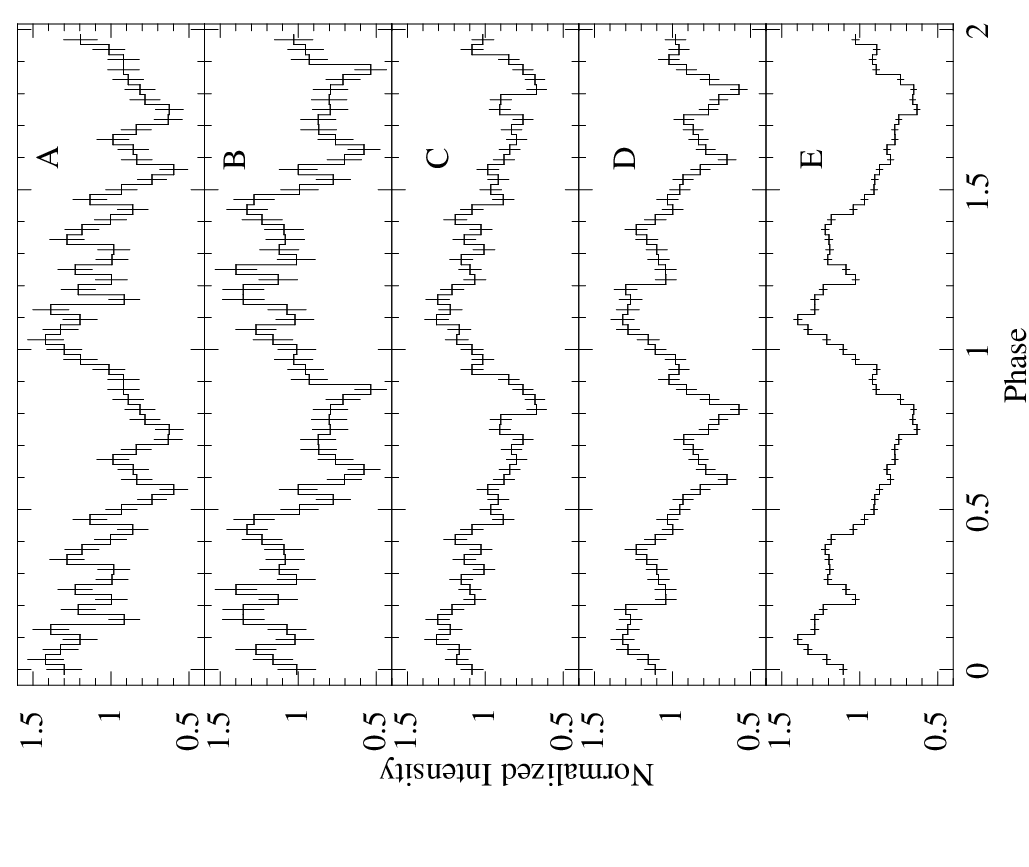}
\caption{\rmfamily{The left- and right-hand panels show the time resolved pulse profiles for 
XIS (0.5-12 {\rm{keV}}) and PIN (10-70 {\rm{keV})}, respectively, folded
with a period of 36.930 \rm{s} and a period derivative of  $3.1847\times{10^{-8}}$ \rm{s s$^{-1}$} at
epoch MJD 55830.}}
\label{pp_tr}
\end{figure*}

\begin{figure*}
\begin{center}
\includegraphics[width=8cm,height=7.5cm,angle=-90]{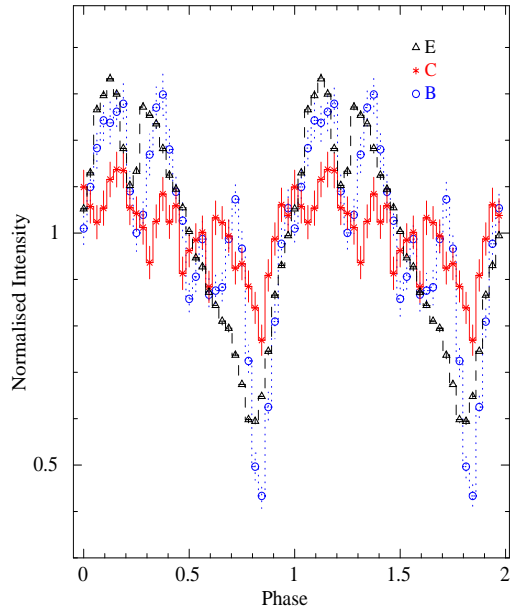}
\end{center}
\caption{\rmfamily{XIS pulse profiles folded with 
the same period and period derivative for the same epoch as Fig. \ref{pp_tr} for segments B, C and E are overlaid. The pulse fraction change in C, and the shift in phase is apparent compared to B and E.}}
\label{1pp_tr_overlayed} 
\end{figure*}

\begin{figure*}
\includegraphics[scale=0.6,angle=-90]{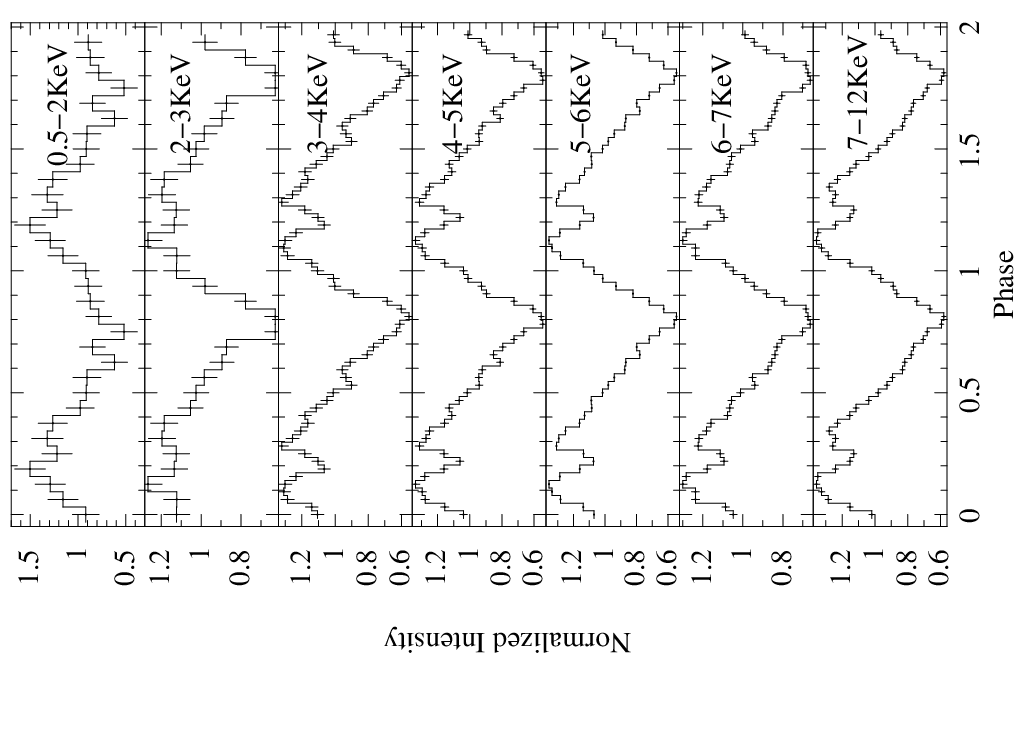}
\includegraphics[scale=0.6,angle=-90]{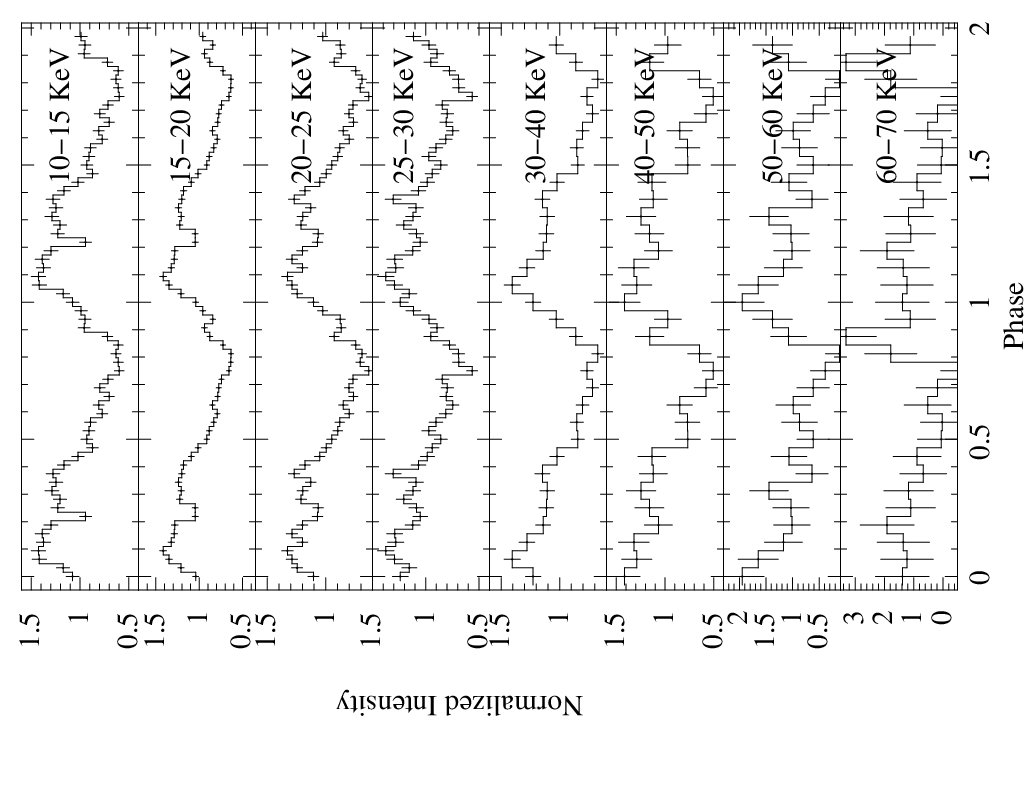}
\caption{\rmfamily{The left- and right- hand panels show the energy resolved pulse profiles for segment E for XIS and PIN respectively folded
with a period of 36.930 \rm{s} and a period derivative of  $3.1847\times{10^{-8}}$ \rm{s s$^{-1}$} at 
epoch MJD 55830. The pulse profiles do not show dependence on energy.}}
\label{pp_er}
\end{figure*}

A time-resolved study of the pulse profiles was carried out for both the XIS and
PIN data. 
Instead of correcting for the pulse arrival time delays due to the orbital motion of the neutron star, in
each of the segments we have allowed for a period derivative while applying the
epoch folding technique to measure the corresponding pulse periods. 
To choose the period derivative, we have divided the light curve into 10 segments and noted the period in each segment. The difference in the
periods divided by the total duration of the light curve gives an approximate value of period derivative. 
We further refined the period derivative by carrying out the pulse period determination (with the tool `efsearch' of 
\texttt{FTOOLS} ) 
repeatedly with different trial period derivatives. Corresponding to the maximum $\chi^{2}$  value obtained, 
the period derivative was found to be $3.1847\times{10^{-8}}$ \rm{s s$^{-1}$}
and the pulse period was 36.930($\pm$ 0.002) \rm{s} at MJD 55830. 
These values
were used to create the normalized profiles i.e., folded light curves, normalized by dividing by the average source intensity in 
each frame which are shown in Fig. \ref{pp_tr}. 
The pulse profiles for segments A and B do not show any significant difference. 
However, segment C shows a smaller pulse fraction compared to the rest.
Also, there is a slight phase shift in C compared to B and D.
The broad features of pulse profiles of the segments D and segment E are
identical but some of the narrow features seen in segment D are not see in
segment E, the latter segment having a higher count rate.
For a better comparison of the pulse profile changes, we have overlaid the
XIS pulse profiles for three segments in Fig. \ref{1pp_tr_overlayed}.  
We have created energy resolved pulse profiles only for the segment E which
has the highest count rate shown in Fig. \ref{pp_er}. It shows that the profiles do not show any
significant energy dependence and pulsations are detected up to 70 \rm{keV}.
\section{Spectral Analysis}
\label{sect:spectral analysis}
\begin{figure*}
\includegraphics[height=4.2in,width=3.3in,angle=-90]{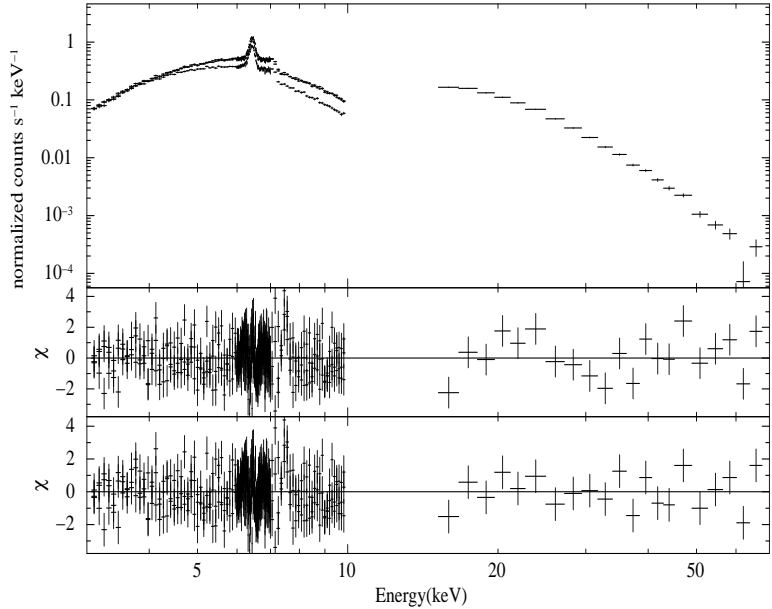}
\caption{\rmfamily{Time-averaged spectrum for OAO 1657--415 using the NPEX model. The middle panel shows the residue without using 
CYCLABS and the lowermost panel
shows the residue when CYCLABS is included.}}
\label{npex_tot1}
\end{figure*}

We performed time averaged spectral analysis of OAO 1657--415  using spectra
from all the three XISs and the PIN.
Spectral fitting was performed using \texttt{XSPEC} v12.7.1.
Artificial structures are known in the XIS spectra around the Si
edge and Au edge and the energy range of 1.75-2.23 \rmfamily{keV} is usually
not used for spectral fitting. Additionally, owing to strong absorption,
the spectrum of OAO 1657--415 has very limited statistics below 3 \rmfamily{keV}.
For spectral fit we have therefore, chosen the energy range of 3-10 \rmfamily{keV}
for the XISs and 15-70 \rmfamily{keV} for the PIN spectrum respectively.
We fitted the 
spectra simultaneously with all parameters tied, except the relative 
instrument normalizations which were kept free. The 2048 channel XIS spectra were rebinned by a factor of 10 up to 5 \rmfamily{keV}, by 2 from 5 to 7 \rmfamily{keV} and by 14 for the rest. 
The PIN spectra were binned by a factor of 4 till 22 \rmfamily{keV}, by 6
from 22 to 45 \rmfamily{keV}, and by 10 for the rest. 
To fit the continuum for the spectra, we tried using the standard continuum models\footnote{
\url{http://heasarc.gsfc.nasa.gov/xanadu/xspec/manual/XspecModels.html}} used for 
HMXBs like  HIGHECUT \citep{W83, CO01}, NPEX \citep{M95},
COMPTT \citep{TI94} and FDCUT \citep{T86}. 
Good fit for the average spectra was obtained with two of the continuum models,
NPEX and HIGHECUT. In the NPEX model, the photon index of the 
second power-law component was fixed at 2.0. 
For both, a partial covering absorption and two Gaussian emission line components were required to account for the two lines at 6.4 keV and 7.1 keV, 
which correspond to $K{\alpha}$ and $K{\beta}$ fluorescence emission lines of iron. 
For the time-averaged spectra showed some dip like residuals in the PIN data. 
We used cyclotron absorption profiles GABS and CYCLABS
 to account for the residuals, 
of which CYCLABS provided a better fit.
However, all parameters of the cyclotron line could not be constrained and we have
fixed the width of the line to 2 \rm{keV}. Inclusion of CYCLABS along with NPEX resulted in a decrease in $\chi^{2}$ from 628 (415 degrees of freedom) to 
618 (413 degrees of freedom).
The dip like feature for HIGHECUT and NPEX were seen at 34 and 32 \rm{keV}, respectively.
We note here that possible presence of a cyclotron line in OAO 1657--415
at a similar energy was reported earlier from \emph{Beppo}-SAX observations
\citep{O99}. \\
Fig. \ref{npex_tot1} shows the average spectra fitted with NPEX model and the spectral parameters for the best fit obtained for the two models are given in Table
\ref{spec_par}. \\
To calculate the significance of emission line features which are additive components in \texttt{XSPEC} like the iron line emission, the \emph{F}-test routine in 
\texttt{XSPEC} package 
is best suited to perform the significance test barring precautions as mentioned in \cite{P02}. However, for the purpose of detecting the significance of multiplicative 
components like the cyclotron line, the same
is not valid. Hence, we use the \emph{F}-test routine available in \texttt{IDL} package MPFTEST\footnote{\url{http://www.physics.wisc.edu/~craigm/idl/down/mpftest.pro}}  for significance test of 
cyclotron line \citep{D13}. The probability of chance improvement (PCI) is evaluated for the NPEX model used to fit the spectrum with 
and without cyclotron line. The estimated PCI values after addition of cyclotron line component to the NPEX is $\sim$ 45 \% . 
Therefore, a cyclotron line at around 36 {\rm{keV}} which was detected with \emph{Beppo}-SAX
is not confirmed in the \emph{Suzaku} spectrum.

\subsection{Time-resolved Spectroscopy}
\begin{figure*}
\begin{center}$
 \begin{array}{cc}
\includegraphics[height=2.8in, width=2in, angle=-90]{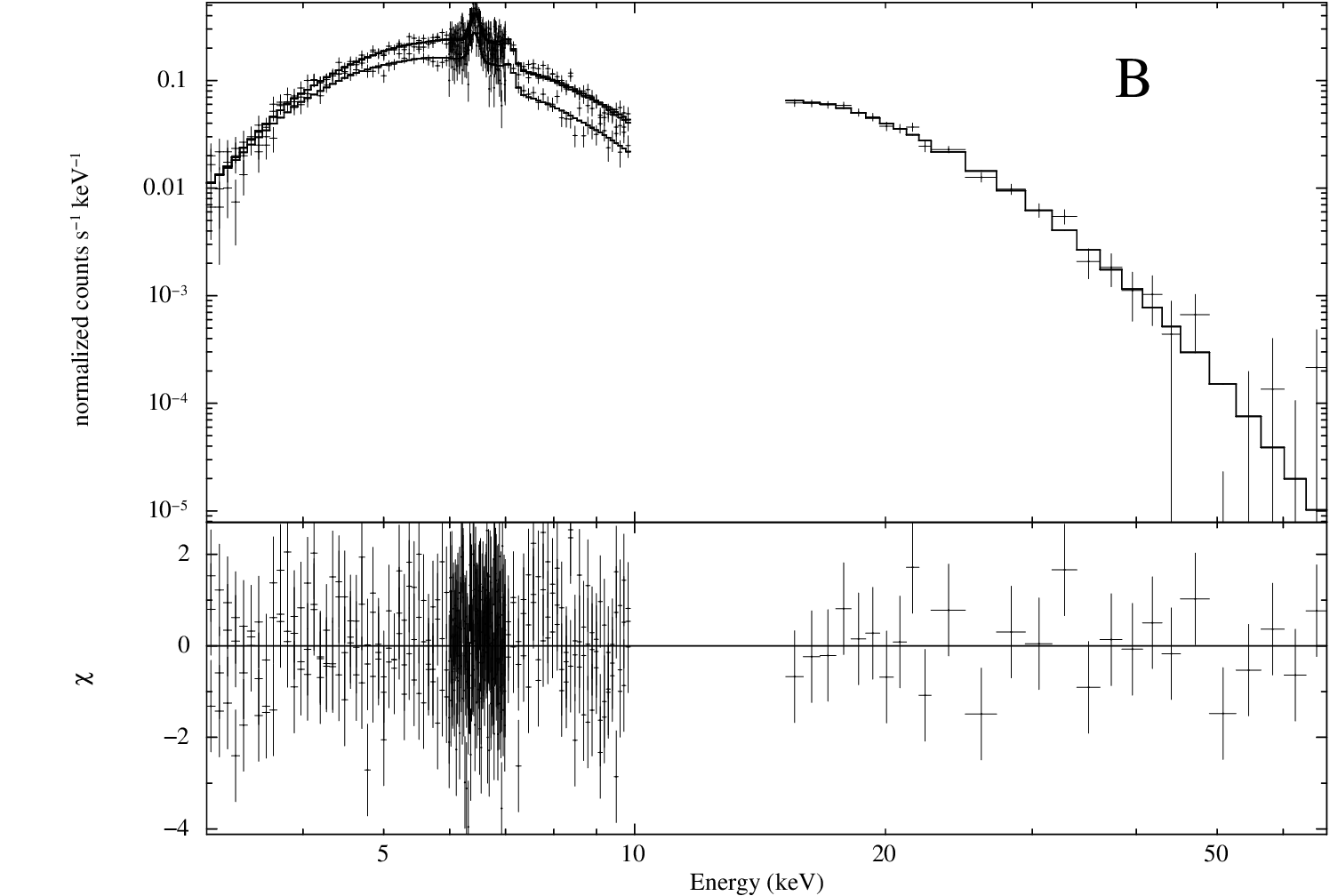} &
\includegraphics[height=2.8in, width=2.4in, angle=-90]{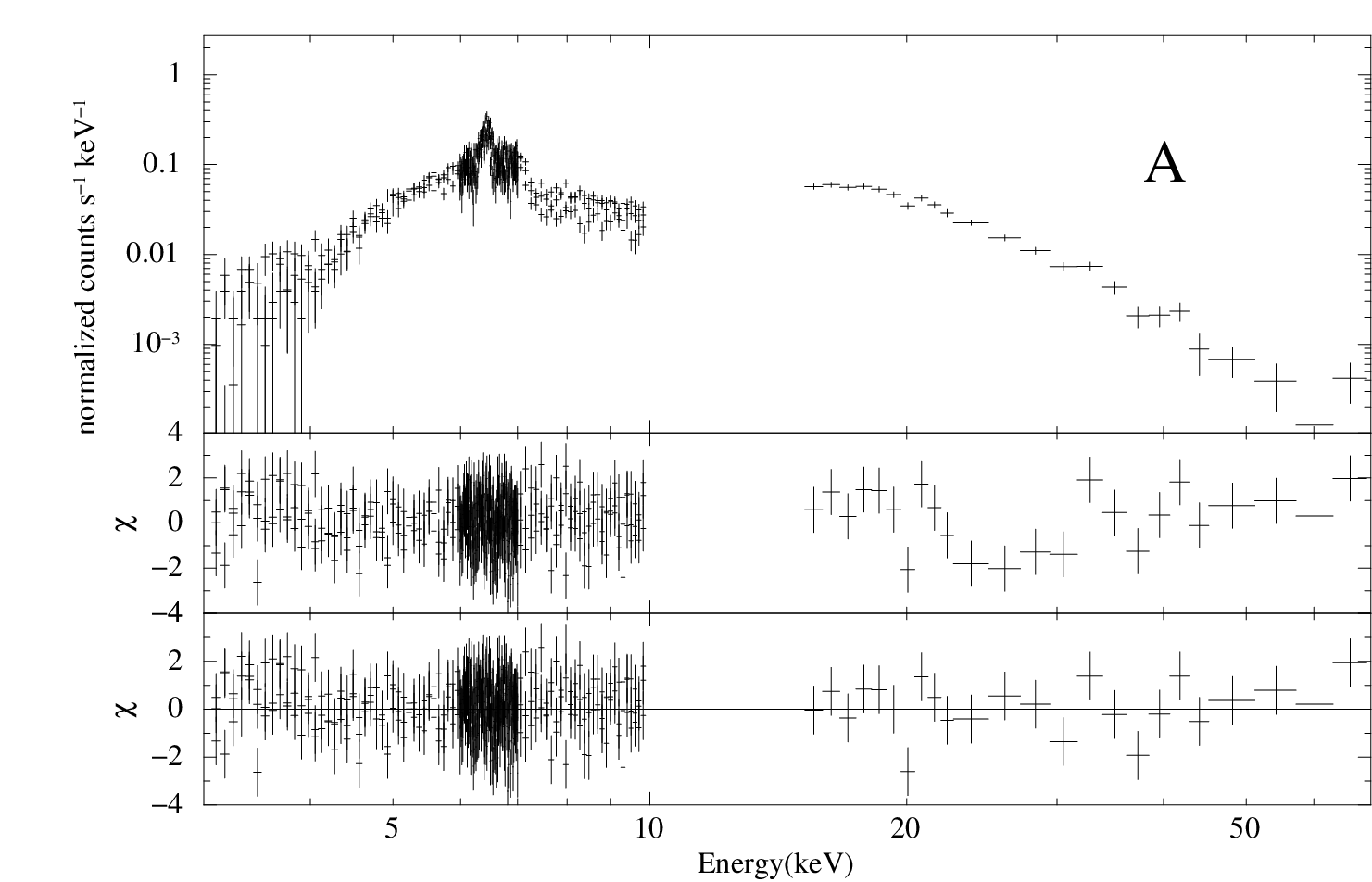} \\
\includegraphics[height=2.8in, width=1.8in, angle=-90]{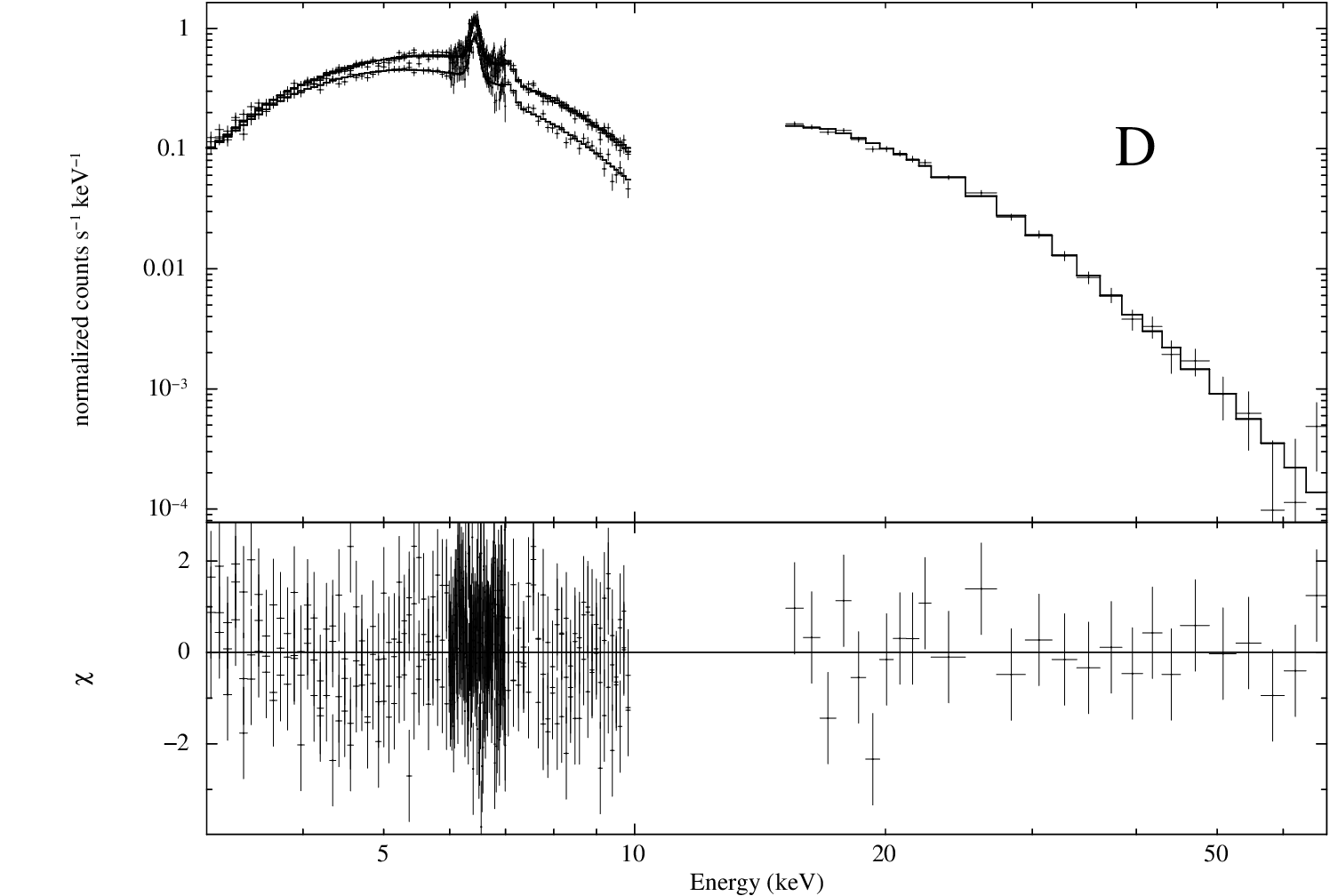} &
\includegraphics[height=2.8in, width=2.4in, angle=-90]{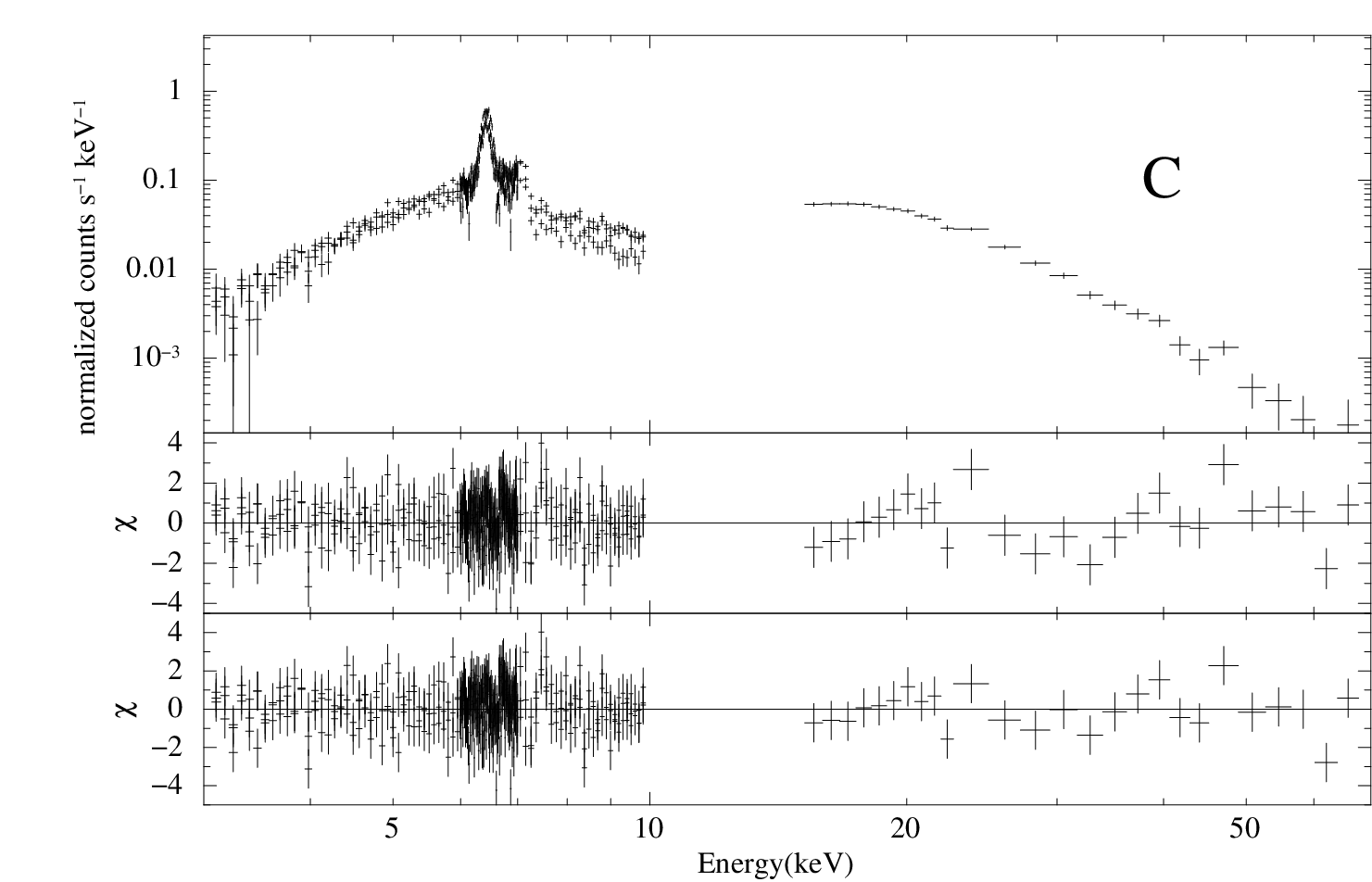} \\
\includegraphics[height=2.8in, width=1.8in, angle=-90]{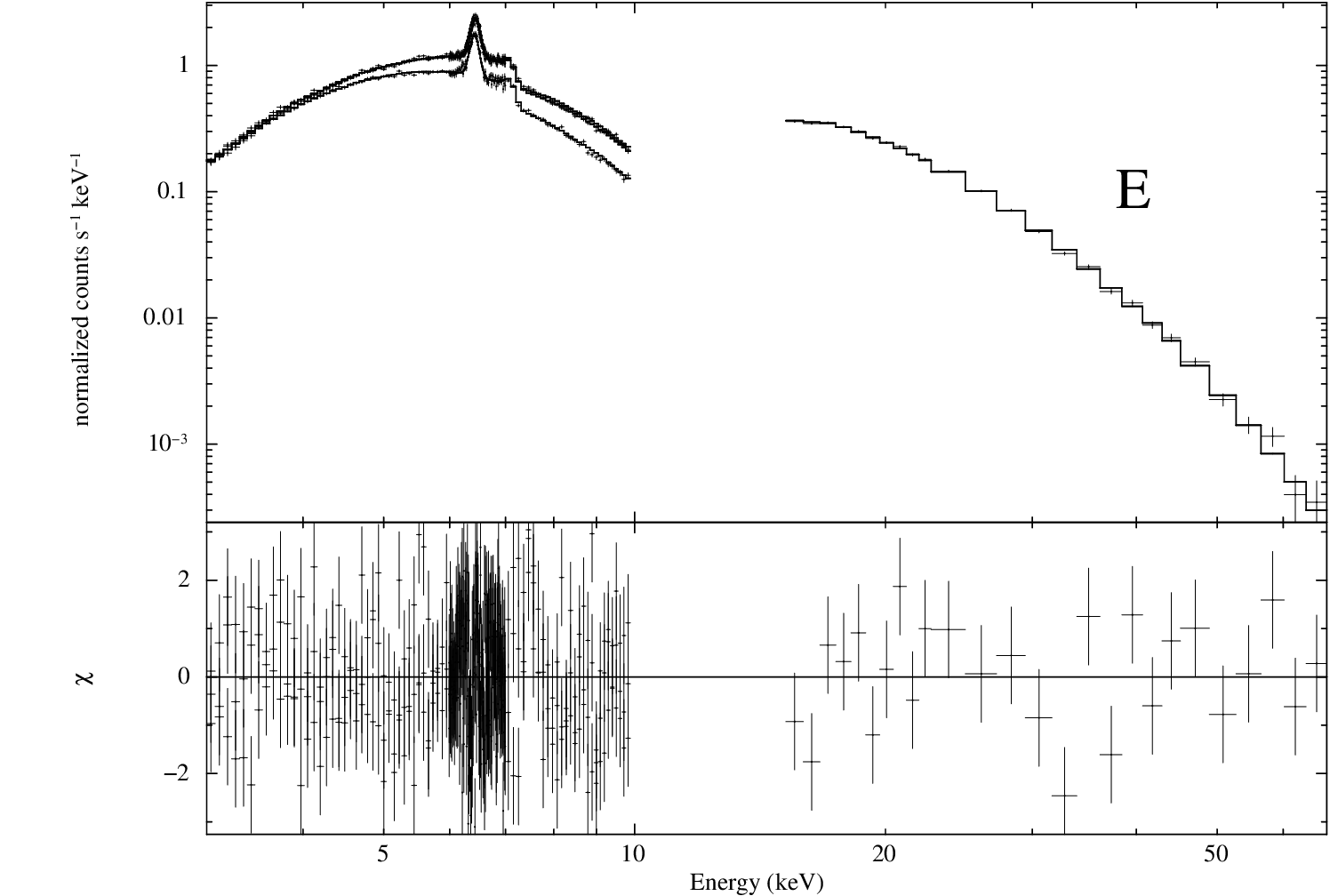} 
\end{array}$
\end{center}
\caption{Time-resolved spectrum for OAO 1657--415 using the NPEX model. The middle panel for segments A and C shows the residue obtained after fitting without
using CYCLABS and the lowermost panel in both represent the residuals after CYCLABS is fitted.}
\label{tr_spec}
\end{figure*}

\begin{figure*}
\centering
\includegraphics[scale=0.55,angle=0]{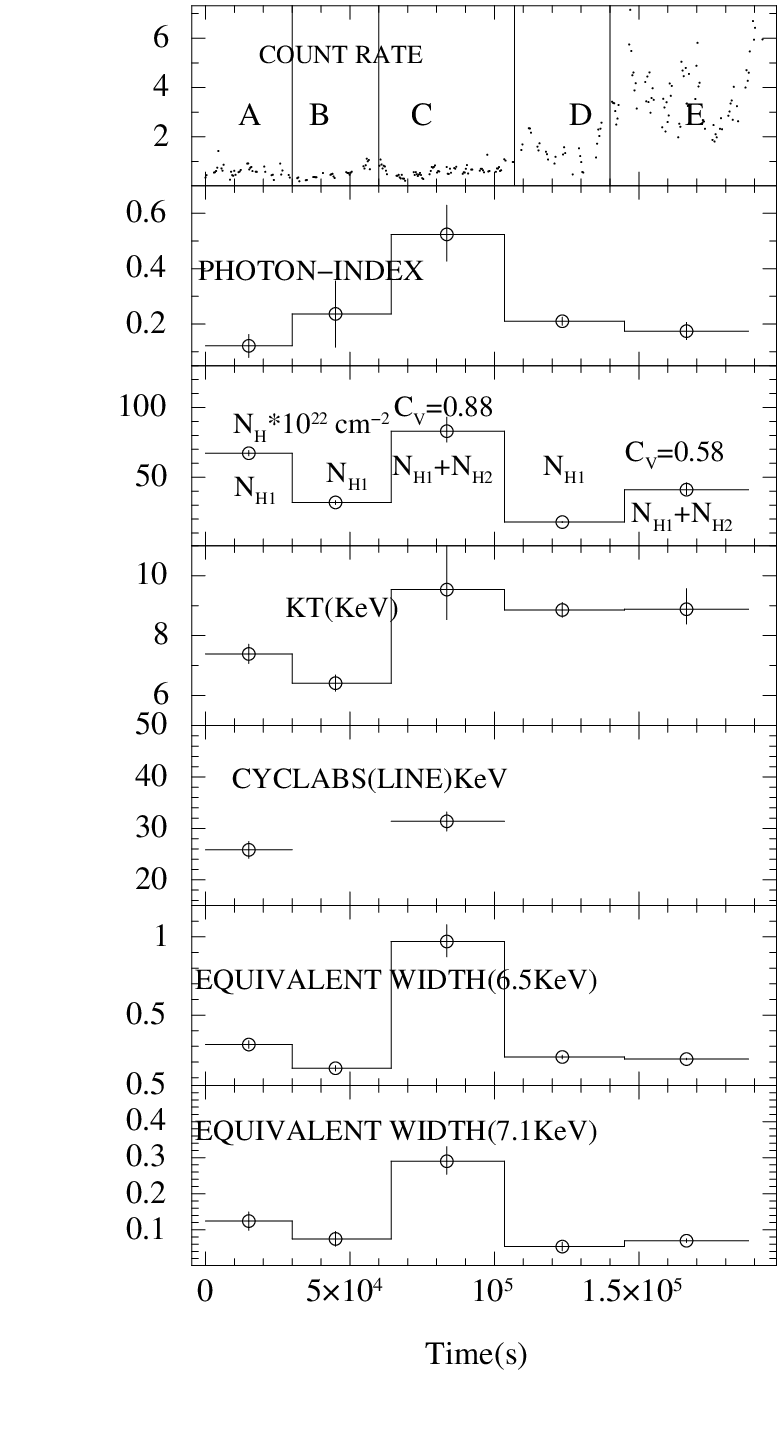}
\includegraphics[scale=0.55,angle=0]{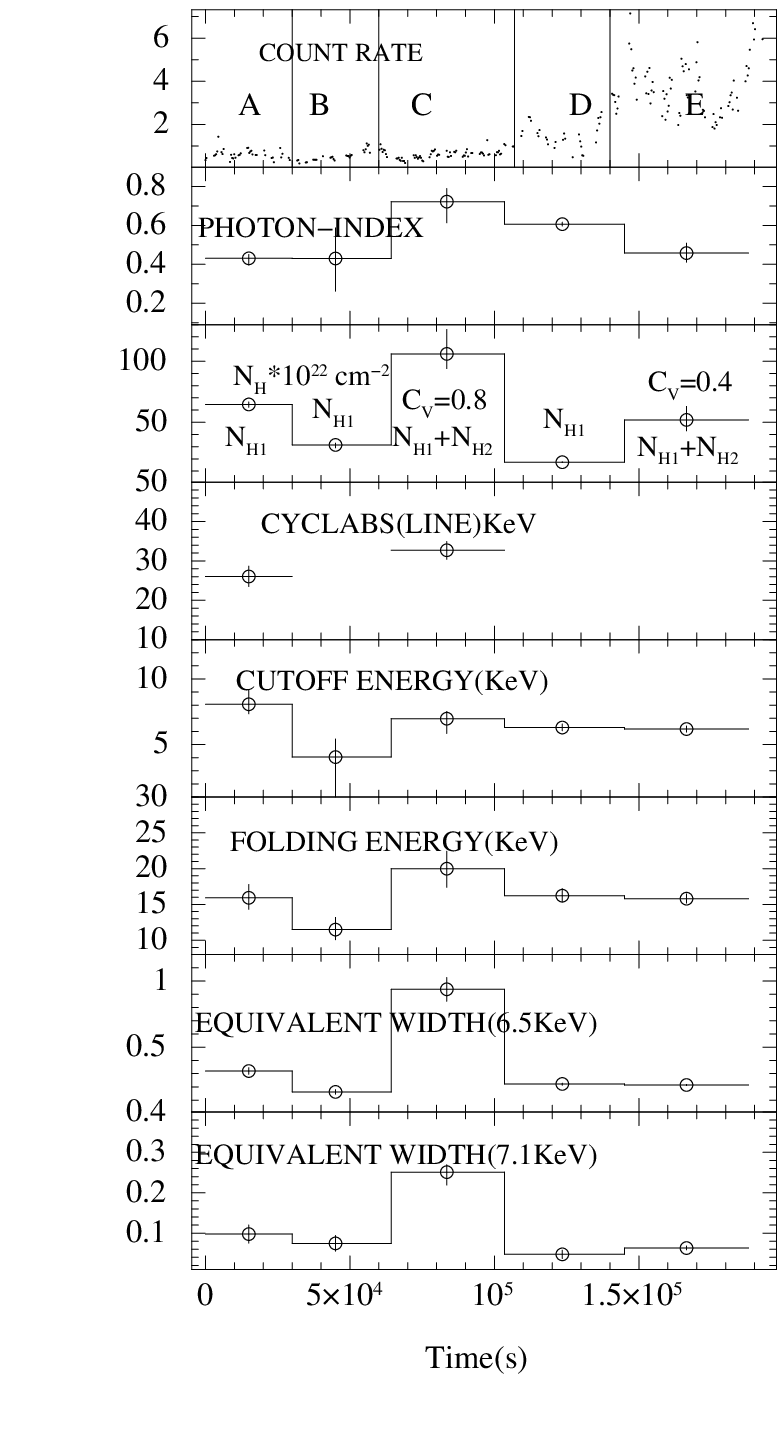}
\caption{\rmfamily{Variation of spectral parameters with time using NPEX is shown in the left and to the right is the variation of 
spectral parameters using HIGHECUT model. The topmost panel in both figures show count rate.}}
\label{npex_he_param_cyclabs}
\end{figure*}

\begin{figure*}
\centering
\includegraphics[scale=0.4,angle=-90]{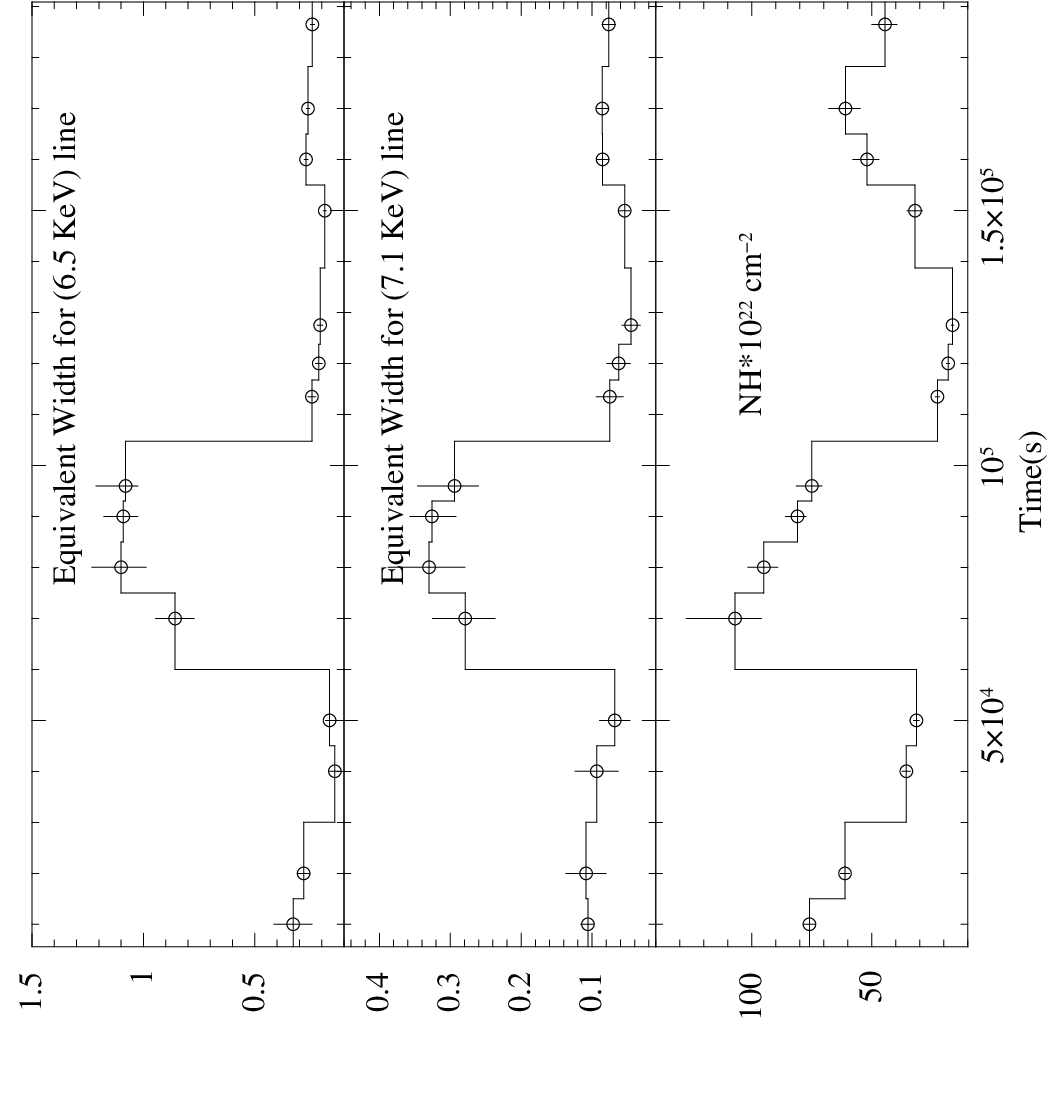}
\includegraphics[scale=0.4,angle=-90]{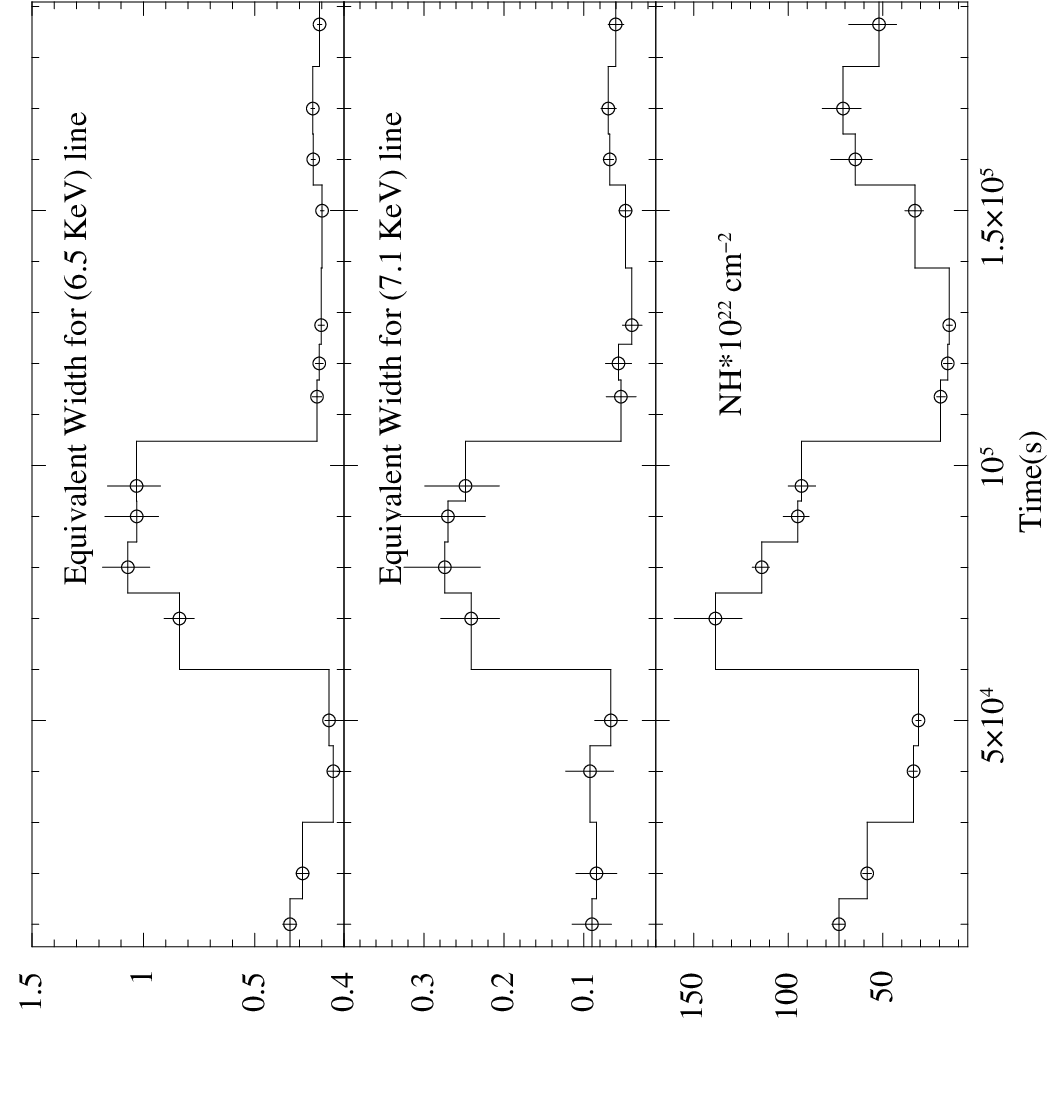}
\caption{\rmfamily{Variation of equivalent width for Gaussian line elements at 6.5 and 7.1 \rm{keV} for the NPEX model to the left and 
HIGHECUT to the right.}}
\label{eqw12_nh_npex_he}
\end{figure*}

The variable hardness ratio shown in Fig. \ref{lcurve} indicates significant spectral changes during the duration of the observation.
To probe into the details of the spectral variations, we divided the light curve into five segments on the basis of their hardness ratio as shown in 
Fig. \ref{lcurve}. 
The time-resolved spectrum was fitted with the same model as used for time-averaged spectra. 
However, a partial covering absorption was not required for the segments A, B and D.
Also, the negative residuals around 33 \rm{keV} in the PIN data mentioned in the previous section and shown in
Fig. \ref{tr_spec} was visible only in the two segments A and C.
The variations of the spectral parameters with time obtained from spectral fits with the two continuum models
NPEX and HIGHECUT are shown in Fig. \ref{npex_he_param_cyclabs}. Variations of the spectral parameters obtained with the two models
are consistent indicating the robustness of the results. 
As seen in Fig. \ref{npex_he_param_cyclabs}, the equivalent widths for the two iron line elements increase largely in the third segment C. Segments A and C have similar
hardness ratios. However, the equivalent widths and $N_{\rm H}$ in segment A is smaller compared to segment C. 
This difference led us to check the variation of equivalent widths and $N_{\rm H}$ with time in finer time intervals. This variation is shown in Fig. \ref{eqw12_nh_npex_he}.
We shall discuss about this further in the next section.  

\begin{table*} 
\scriptsize
\caption{Best-fitting parameters of the time--averaged spectra for OAO 1657--415  during the \emph{Suzaku} observation.
 Errors quoted are for 90 per cent confidence range.}
\label{table}
\centering
\begin{tabular}{|c | c | c |c |c |c |c |c |}
\hline
Parameter & NPEX & HIGHECUT \\
\hline
$N_{\rm H1}$ ($10^{22}$ atoms \rm{cm$^{-2}$}) & $15.6_{-0.9}^{+0.8}$ &   $16.7_{-0.65}^{+0.71}$ \\
$N_{\rm H2}$ ($10^{22}$ atoms \rm{cm$^{-2}$}) & $34_{-3.1}^{+3.5}$ &   $42.7_{-5.8}^{+9.1}$ \\	
PowIndex & $0.22$*$_{-0.01}^{+0.02}$  & $0.52_{-0.02}^{+0.01}$ \\
CvrFract & $0.58_{-0.04}^{+0.05}$ & $0.46_{-0.03}^{+0.03}$ \\
\rm{E}$_{highecut}$ (keV) & $8.6_{-0.14}^{+0.26}$ & -- \\
Ecut (keV) &  -- & $6.2_{-0.17}^{+0.16}$ \\
Efold (keV) & -- & $16.6_{-0.6}^{+0.4}$ \\
Depth(keV)& $0.11_{-0.05}^{+0.05}$ & $0.1_{-0.04}^{0.04}$ \\
Cyclabs(keV) & $32_{-2.0}^{+2.3}$ &  $34_{-2.4}^{+5.6}$ \\
K$\alpha$ line (keV) & $6.44_{-0.003}^{+0.001}$ & $6.43_{-0.002}^{+0.003}$ \\
K$\beta$ line (keV)& $7.09_{-0.01}^{+0.01}$  & $7.11_{-0.006}^{+0.007}$ \\
Equivalent Width for K$\alpha$ line (keV) & $0.261_{-0.006}^{+0.005}$ & $0.261_{0.005}^{+0.006}$ \\
Equivalent Width for K$\beta$ line (keV) & $0.081_{-0.004}^{+0.004}$  & $0.074_{-0.003}^{+0.005}$ \\
 $\chi^{2}_{\nu}$/d.o.f & 1.49/413 (1.51/415 without CYCLABS)  & 1.46/417 (1.47/419 without CYCLABS) \\
\hline
\end{tabular}
\begin{flushleft}
$*$ Photon-index of the second power-law component of the NPEX model is frozen to 2.0, as mentioned in the text.
\end{flushleft}
\label{spec_par}
\end{table*}

\begin{figure*}
\includegraphics[scale=0.5,angle=-90]{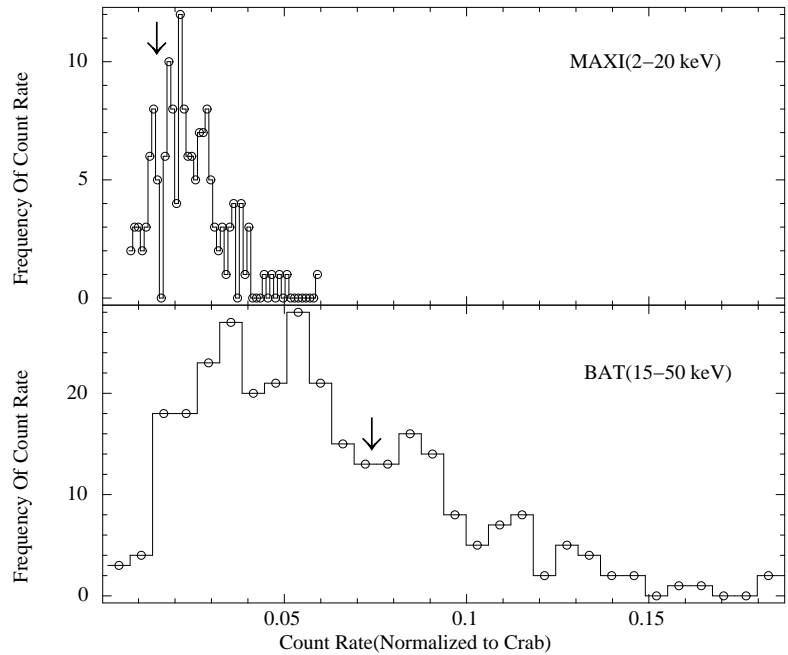}
\caption{\rmfamily{Histogram of the orbit averaged count rate of OAO 1657--415 for \emph{Swift}-BAT and 
\emph{MAXI}-GSC. The count rate (\rm{X}-axis) is normalized to that of
 Crab nebula for the respective instrument. The
arrows represent the orbit corresponding to \emph{Suzaku} observation.}}
\label{combine}
\end{figure*}
\section{Discussion}
\subsection{Flux and spectral variability}
The variation in intensity for this source over a large number of orbits taken 
together has been studied in the past \citep{BS08}.
Fig. \ref{combine} shows the intensity (normalized to Crab) histograms with \emph{MAXI}-GSC\footnote{\url{http://maxi.riken.jp/top/index.php?cid=1&jname=J1700-416}} 
(MJD 55063--56661) and \emph{Swift}-BAT (MJD 53415--56617). 
The orbit averaged intensity corresponding to this \emph{Suzaku} observation are
marked with arrows in the two panel of Fig. \ref{combine} which indicates that in this
orbit OAO 1657--415 did not have extreme property i.e. very high or very low
count rate compared to the long-term average.
In this paper, we discuss the variability of OAO 1657--415 in time-scales that is a fraction of the orbital period.\\
Segment C makes an interesting study. As depicted in Fig. \ref{1pp_tr_overlayed}, we see that pulse fraction suddenly goes very
low in C compared to the other segments.
During this segment, the equivalent width for both Gaussian line elements at 6.4 and 7.1 \rm{keV} 
increases compared to other segments. This change of pulse fraction
in C is more apparent for XIS data than PIN data. This can be due to the
dominance of the Fe line photons, which are unpulsed in nature. 
The variation in the characteristics of the segments are summarized below:\\ 
 \begin{enumerate}
 \item Segment A: 
 large hardness ratio with large $N_{\rm H}$ and moderate equivalent width.
 \item Segment B: 
 hardness ratio is low with low $N_{\rm H}$ and low equivalent width.
 \item Segment C: 
 hardness ratio is high with very large $N_{\rm H}$ and large equivalent width.
 \item Segment D: 
 same as B. 
\item Segment E: 
 same as B, but flux higher by a factor of $\sim$ 6.
\end{enumerate}
One possible explanation for the above observations could be as follows:
during segment A, the pulsar may be passing behind a dense clump of matter. This causes absorption of soft X-rays, leading to an increase in hardness ratio and 
large value of $N_{\rm H}$ but low equivalent width if the clump subtends a small solid angle to the source. If the neutron star had been passing through the clump,
a large equivalent width would be seen.\\
During segment B, the hardness ratio, $N_{\rm H}$ and equivalent width are low. Here the pulsar may be passing
through a region where there is very scanty material.\\
In segment C however, a large increase in hardness ratio, $N_{\rm H}$, and equivalent width indicates that instead of
being behind a cloud, the neutron star is now passing through a dense clump of matter. The large soft X-ray
absorption leads to increase in the measured hardness ratio and $N_{\rm H}$, while a 4$\pi$ solid angle of the clump causes
a large equivalent width of the iron fluorescence lines. This should also lead to an increase in the mass accretion rate
on to the neutron star and its X-ray luminosity, probably with a delay corresponding to the viscous time-scale from the
material capture radius in the accretion disc to the neutron star.\\
In segment D, the hardness ratio, $N_{\rm H}$ and equivalent width are all low (same scenario as segment B). \\
In segment E also, the hardness ratio, $N_{\rm H}$ and equivalent width are low. However, there is an increase in luminosity (starting from segment D that lasts 
through segment E). 
This could be the increase in luminosity resulting from the increased mass capture from the clump that the
neutron star encountered in segment C. The time lag between the increase mass accretion in segment C, and the increase
in X-ray luminosity which is manifested in segments D and E is less than one day. \\
The viscous time-scale depends on the specific angular momentum of the captured material with respect to the neutron star. In the case of the
 HMXB GX 301-2, which has an orbital period
of 41.5 \rm{d} and which shows a large flare every orbit at the orbital phase of 0.95, an enhanced accretion at orbital phase 0.92($\sim$ 1.2 \rm{d} earlier than the flare) was shown to reproduce 
the flare \citep{PR01}.

\subsection{Comparison with SFXTs and supergiant systems}
The spectral and intensity variations of the Suzaku observation in this source shows that the stellar wind emitted from 
the companion star is inhomogeneous with many clumps as indicated in \cite{OF12}.
A rough estimate about the size of the clump which the neutron star passes through during segment C
is made as follows:\\
The time of passage of the neutron star through the segment C is $\sim$ 47 \rm{ks}. 
Assuming the relative velocity of the wind $v_\mathrm{rel} =300\ \mathrm{km\ s^{-1}}$ at the neutron star, we obtain the clump radius of segment C,  $R_c \simeq v_\mathrm{rel}t_f/2$ = $7\times10^{11}$ cm. 
For the C segment, the column density of $N_{\rm H}$ is $\sim$ 10$^{24}$ \rm{cm$^{-2}$}. \\ 
Hence, the mass of clump is
     $\sim 3\times10^{24}\rm{g}$ .\\
In similar lines, \cite{BO11} discussed that X-ray flares observed from an SFXT, 
IGR J18410--0535 as being due to accretion of matter from a massive clump on to the neutron star, the mass of
the clump being $\sim 1.4\times10^{22}\rm{g}$.  
 \\
Here, it is worthwhile
to note the similarity of OAO 1657--415 with Vela X-1. Vela X-1 is embedded in the dense
stellar wind of its optical companion \citep{NA86} and displays a strong time variability \cite[]{KR08, S08}. 
Owing to the high X-ray variability of both OAO 1657--415 and Vela X-1, 
they can be seen as a class of systems which could
represent a possible link between SFXTs and normal HMXBs. \\
SFXTs show irregular outbursts, lasting from minutes to hours, with peak X-ray luminosities between $10^{36}$ and
$10^{37} erg s^{-1}$ in contrast to a quiescent phase when the typical luminosities maybe $10^{32} erg s^{-1}$ \citep{GL04,SL05,GS05,LV05,MS06,GO07}.
These variations could be due to the clumpiness of stellar wind which may lead to variations of the 
density and velocity of matter around the neutron star, resulting in the fluctuation of the accretion rate \cite[]{KR08,DU09} or when accretion is regulated by magnetospheric barrier \citep{BO08}. Though
the study of variability of OAO 1657--415 here is in longer time-scales compared to SFXTs, further study of
this class of objects will help us in better understanding of the physical origin of the X-ray 
variability and provide a link between SFXTs and supergiant HMXBs.

\section{Acknowledgements}
The data for this work have been obtained through the High Energy Astrophysics Science Archive (HEASARC) Online Service provided by NASA/GSFC. 
We have also made use of public light curves from \emph{Swift} and \emph{MAXI} site. We would also like to thank the anonymous referee and Sachindra Naik 
for invaluable comments and suggestions.

\bibliography{draft2}{}

\begin{thebibliography}{38}
\expandafter\ifx\csname natexlab\endcsname\relax\def\natexlab#1{#1}\fi

\bibitem[{{Audley} {et~al}\mbox{.}(2006){Audley}, {Nagase}, {Mitsuda},
  {Angelini}, \& {Kelley}}]{A06}
{Audley} M.~D., {Nagase} F., {Mitsuda} K., {Angelini} L., {Kelley} R.~L., 2006,
  \mnras, 367, 1147

\bibitem[{{Barnstedt} {et~al}\mbox{.}(2008){Barnstedt}, {Staubert},
  {Santangelo}, {Ferrigno}, {Horns}, {Klochkov}, {Kretschmar}, {Kreykenbohm},
  {Segreto}, \& {Wilms}}]{BS08}
{Barnstedt} J. {et~al.}, 2008, \aap, 486, 293

\bibitem[{{Baykal}(1997)}]{B97}
{Baykal} A., 1997, \aap, 319, 515

\bibitem[{{Baykal}(2000)}]{B00}
{Baykal} A., 2000, \mnras, 313, 637

\bibitem[{{Bildsten} {et~al}\mbox{.}(1997){Bildsten}, {Chakrabarty}, {Chiu},
  {Finger}, {Koh}, {Nelson}, {Prince}, {Rubin}, {Scott}, {Stollberg},
  {Vaughan}, {Wilson}, \& {Wilson}}]{BI97}
{Bildsten} L. {et~al.}, 1997, \apjs, 113, 367

\bibitem[{{Bozzo}, {Falanga} \& {Stella}(2008){Bozzo}, {Falanga}, \&
  {Stella}}]{BO08}
{Bozzo} E., {Falanga} M., {Stella} L., 2008, \apj, 683, 1031

\bibitem[{{Bozzo} {et~al}\mbox{.}(2011){Bozzo}, {Giunta}, {Cusumano},
  {Ferrigno}, {Walter}, {Campana}, {Falanga}, {Israel}, \& {Stella}}]{BO11}
{Bozzo} E. {et~al.}, 2011, \aap, 531, A130

\bibitem[{{Chakrabarty} {et~al}\mbox{.}(1993){Chakrabarty}, {Grunsfeld},
  {Prince}, {Bildsten}, {Finger}, {Wilson}, {Fishman}, {Meegan}, \&
  {Paciesas}}]{C93}
{Chakrabarty} D. {et~al.}, 1993, \apjl, 403, L33

\bibitem[{{Chakrabarty} {et~al}\mbox{.}(2002){Chakrabarty}, {Wang}, {Juett},
  {Lee}, \& {Roche}}]{C02}
{Chakrabarty} D., {Wang} Z., {Juett} A.~M., {Lee} J.~C., {Roche} P., 2002,
  \apj, 573, 789

\bibitem[{{Corbet}(1986)}]{C86}
{Corbet} R.~H.~D., 1986, \mnras, 220, 1047

\bibitem[{{DeCesar} {et~al}\mbox{.}(2013){DeCesar}, {Boyd}, {Pottschmidt},
  {Wilms}, {Suchy}, \& {Miller}}]{D13}
{DeCesar} M.~E., {Boyd} P.~T., {Pottschmidt} K., {Wilms} J., {Suchy} S.,
  {Miller} M.~C., 2013, \apj, 762, 61

\bibitem[{{Ducci} {et~al}\mbox{.}(2009){Ducci}, {Sidoli}, {Mereghetti},
  {Paizis}, \& {Romano}}]{DU09}
{Ducci} L., {Sidoli} L., {Mereghetti} S., {Paizis} A., {Romano} P., 2009,
  \mnras, 398, 2152

\bibitem[{{Fukazawa} {et~al}\mbox{.}(2009){Fukazawa}, {Mizuno}, {Watanabe},
  {Kokubun}, {Takahashi}, {Kawano}, {Nishino}, {Sasada}, {Shirai}, {Takahashi},
  {Umeki}, {Yamasaki}, {Yasuda}, {Bamba}, {Ohno}, {Takahashi}, {Ushio},
  {Enoto}, {Kitaguchi}, {Makishima}, {Nakazawa}, {Uehara}, {Yamada}, {Yuasa},
  {Isobe}, {Kawaharada}, {Tanaka}, {Tashiro}, {Terada}, \& {Yamaoka}}]{F09}
{Fukazawa} Y. {et~al.}, 2009, \pasj, 61, 17

\bibitem[{{Gonz{\'a}lez-Riestra} {et~al}\mbox{.}(2004){Gonz{\'a}lez-Riestra},
  {Oosterbroek}, {Kuulkers}, {Orr}, \& {Parmar}}]{GL04}
{Gonz{\'a}lez-Riestra} R., {Oosterbroek} T., {Kuulkers} E., {Orr} A., {Parmar}
  A.~N., 2004, \aap, 420, 589

\bibitem[{{G{\"o}tz} {et~al}\mbox{.}(2007){G{\"o}tz}, {Falanga}, {Senziani},
  {De Luca}, {Schanne}, \& {von Kienlin}}]{GO07}
{G{\"o}tz} D., {Falanga} M., {Senziani} F., {De Luca} A., {Schanne} S., {von
  Kienlin} A., 2007, \apjl, 655, L101

\bibitem[{{Grebenev} \& {Sunyaev}(2005)}]{GS05}
{Grebenev} S.~A., {Sunyaev} R.~A., 2005, Astronomy Letters, 31, 672

\bibitem[{{Heindl} {et~al}\mbox{.}(2001){Heindl}, {Coburn}, {Gruber},
  {Rothschild}, {Kreykenbohm}, {Wilms}, \& {Staubert}}]{CO01}
{Heindl} W.~A., {Coburn} W., {Gruber} D.~E., {Rothschild} R.~E., {Kreykenbohm}
  I., {Wilms} J., {Staubert} R., 2001, \apjl, 563, L35

\bibitem[{{Jenke} {et~al}\mbox{.}(2012){Jenke}, {Finger}, {Wilson-Hodge}, \&
  {Camero-Arranz}}]{J12}
{Jenke} P.~A., {Finger} M.~H., {Wilson-Hodge} C.~A., {Camero-Arranz} A., 2012,
  \apj, 759, 124

\bibitem[{{Koyama} {et~al}\mbox{.}(2007){Koyama}, {Tsunemi}, {Dotani}, {Bautz},
  {Hayashida}, {Tsuru}, {Matsumoto}, {Ogawara}, {Ricker}, {Doty}, {Kissel},
  {Foster}, {Nakajima}, {Yamaguchi}, {Mori}, {Sakano}, {Hamaguchi},
  {Nishiuchi}, {Miyata}, {Torii}, {Namiki}, {Katsuda}, {Matsuura}, {Miyauchi},
  {Anabuki}, {Tawa}, {Ozaki}, {Murakami}, {Maeda}, {Ichikawa}, {Prigozhin},
  {Boughan}, {Lamarr}, {Miller}, {Burke}, {Gregory}, {Pillsbury}, {Bamba},
  {Hiraga}, {Senda}, {Katayama}, {Kitamoto}, {Tsujimoto}, {Kohmura}, {Tsuboi},
  \& {Awaki}}]{K07}
{Koyama} K. {et~al.}, 2007, \pasj, 59, 23

\bibitem[{{Kreykenbohm} {et~al}\mbox{.}(2008){Kreykenbohm}, {Wilms},
  {Kretschmar}, {Torrej{\'o}n}, {Pottschmidt}, {Hanke}, {Santangelo},
  {Ferrigno}, \& {Staubert}}]{KR08}
{Kreykenbohm} I. {et~al.}, 2008, \aap, 492, 511

\bibitem[{{Lutovinov} {et~al}\mbox{.}(2005){Lutovinov}, {Revnivtsev},
  {Gilfanov}, {Shtykovskiy}, {Molkov}, \& {Sunyaev}}]{LV05}
{Lutovinov} A., {Revnivtsev} M., {Gilfanov} M., {Shtykovskiy} P., {Molkov} S.,
  {Sunyaev} R., 2005, \aap, 444, 821

\bibitem[{{Masetti} {et~al}\mbox{.}(2006){Masetti}, {Pretorius}, {Palazzi},
  {Bassani}, {Bazzano}, {Bird}, {Charles}, {Dean}, {Malizia}, {Nkundabakura},
  {Stephen}, \& {Ubertini}}]{MS06}
{Masetti} N. {et~al.}, 2006, \aap, 449, 1139

\bibitem[{{Mason} {et~al}\mbox{.}(2009){Mason}, {Clark}, {Norton},
  {Negueruela}, \& {Roche}}]{MA09}
{Mason} A.~B., {Clark} J.~S., {Norton} A.~J., {Negueruela} I., {Roche} P.,
  2009, \aap, 505, 281

\bibitem[{{Mihara}(1995)}]{M95}
{Mihara} T., 1995, PhD thesis, , Dept.~of Physics, Univ.~of Tokyo (M95), (1995)

\bibitem[{{Mitsuda} {et~al}\mbox{.}(2007){Mitsuda}, {Bautz}, {Inoue}, {Kelley},
  {Koyama}, {Kunieda}, {Makishima}, {Ogawara}, {Petre}, {Takahashi}, {Tsunemi},
  {White}, {Anabuki}, {Angelini}, {Arnaud}, {Awaki}, {Bamba}, {Boyce}, {Brown},
  {Chan}, {Cottam}, {Dotani}, {Doty}, {Ebisawa}, {Ezoe}, {Fabian}, {Figueroa},
  {Fujimoto}, {Fukazawa}, {Furusho}, {Furuzawa}, {Gendreau}, {Griffiths},
  {Haba}, {Hamaguchi}, {Harrus}, {Hasinger}, {Hatsukade}, {Hayashida}, {Henry},
  {Hiraga}, {Holt}, {Hornschemeier}, {Hughes}, {Hwang}, {Ishida}, {Ishisaki},
  {Isobe}, {Itoh}, {Iyomoto}, {Kahn}, {Kamae}, {Katagiri}, {Kataoka},
  {Katayama}, {Kawai}, {Kilbourne}, {Kinugasa}, {Kissel}, {Kitamoto}, {Kohama},
  {Kohmura}, {Kokubun}, {Kotani}, {Kotoku}, {Kubota}, {Madejski}, {Maeda},
  {Makino}, {Markowitz}, {Matsumoto}, {Matsumoto}, {Matsuoka}, {Matsushita},
  {McCammon}, {Mihara}, {Misaki}, {Miyata}, {Mizuno}, {Mori}, {Mori}, {Morii},
  {Moseley}, {Mukai}, {Murakami}, {Murakami}, {Mushotzky}, {Nagase}, {Namiki},
  {Negoro}, {Nakazawa}, {Nousek}, {Okajima}, {Ogasaka}, {Ohashi}, {Oshima},
  {Ota}, {Ozaki}, {Ozawa}, {Parmar}, {Pence}, {Porter}, {Reeves}, {Ricker},
  {Sakurai}, {Sanders}, {Senda}, {Serlemitsos}, {Shibata}, {Soong}, {Smith},
  {Suzuki}, {Szymkowiak}, {Takahashi}, {Tamagawa}, {Tamura}, {Tamura},
  {Tanaka}, {Tashiro}, {Tawara}, {Terada}, {Terashima}, {Tomida}, {Torii},
  {Tsuboi}, {Tsujimoto}, {Tsuru}, {Turner}, {Ueda}, {Ueno}, {Ueno}, {Uno},
  {Urata}, {Watanabe}, {Yamamoto}, {Yamaoka}, {Yamasaki}, {Yamashita},
  {Yamauchi}, {Yamauchi}, {Yaqoob}, {Yonetoku}, \& {Yoshida}}]{M07}
{Mitsuda} K. {et~al.}, 2007, \pasj, 59, 1

\bibitem[{{Nagase} {et~al}\mbox{.}(1986){Nagase}, {Hayakawa}, {Sato}, {Masai},
  \& {Inoue}}]{NA86}
{Nagase} F., {Hayakawa} S., {Sato} N., {Masai} K., {Inoue} H., 1986, \pasj, 38,
  547

\bibitem[{{Orlandini} {et~al}\mbox{.}(1999){Orlandini}, {dal Fiume}, {del
  Sordo}, {Frontera}, {Parmar}, {Santangelo}, \& {Segreto}}]{O99}
{Orlandini} M., {dal Fiume} D., {del Sordo} S., {Frontera} F., {Parmar} A.~N.,
  {Santangelo} A., {Segreto} A., 1999, \aap, 349, L9

\bibitem[{{Oskinova}, {Feldmeier} \& {Kretschmar}(2013){Oskinova}, {Feldmeier},
  \& {Kretschmar}}]{OF12}
{Oskinova} L.~M., {Feldmeier} A., {Kretschmar} P., 2013, in IAU Symposium, Vol.
  290, IAU Symposium, {Zhang} C.~M., {Belloni} T., {M{\'e}ndez} M., {Zhang}
  S.~N., eds., pp. 287--288

\bibitem[{{Polidan} {et~al}\mbox{.}(1978){Polidan}, {Pollard}, {Sanford}, \&
  {Locke}}]{P78}
{Polidan} R.~S., {Pollard} G.~S.~G., {Sanford} P.~W., {Locke} M.~C., 1978,
  \nat, 275, 296

\bibitem[{{Pravdo} \& {Ghosh}(2001)}]{PR01}
{Pravdo} S.~H., {Ghosh} P., 2001, \apj, 554, 383

\bibitem[{{Protassov} {et~al}\mbox{.}(2002){Protassov}, {van Dyk}, {Connors},
  {Kashyap}, \& {Siemiginowska}}]{P02}
{Protassov} R., {van Dyk} D.~A., {Connors} A., {Kashyap} V.~L., {Siemiginowska}
  A., 2002, \apj, 571, 545

\bibitem[{{Sidoli} {et~al}\mbox{.}(2005){Sidoli}, {Vercellone}, {Mereghetti},
  \& {Tavani}}]{SL05}
{Sidoli} L., {Vercellone} S., {Mereghetti} S., {Tavani} M., 2005, \aap, 429,
  L47

\bibitem[{{Soffitta} {et~al}\mbox{.}(2008){Soffitta}, {Costa}, {Del Monte},
  {Donnarumma}, {Evangelista}, {Feroci}, {Lapshov}, {Lazzarotto}, {Pacciani},
  {Rapisarda}, {Argan}, {Trois}, {Tavani}, {Piano}, {Pucella}, {D'Ammando},
  {Vittorini}, {Bulgarelli}, {Gianotti}, {Trifoglio}, {Di Cocco}, {Labanti},
  {Fuschino}, {Marisaldi}, {Galli}, {Chen}, {Vercellone}, {Giuliani},
  {Mereghetti}, {Perotti}, {Caraveo}, {Pellizzoni}, {Barbiellini}, {Longo},
  {Vallazza}, {Picozza}, {Morselli}, {Prest}, {Lipari}, {Zanello}, {Rappoldi},
  {Pittori}, {Verrecchia}, {Santolamazza}, {Preger}, {Giommi}, \&
  {Salotti}}]{S08}
{Soffitta} P. {et~al.}, 2008, The Astronomer's Telegram, 1782, 1

\bibitem[{{Takahashi} {et~al}\mbox{.}(2007){Takahashi}, {Abe}, {Endo}, {Endo},
  {Ezoe}, {Fukazawa}, {Hamaya}, {Hirakuri}, {Hong}, {Horii}, {Inoue}, {Isobe},
  {Itoh}, {Iyomoto}, {Kamae}, {Kasama}, {Kataoka}, {Kato}, {Kawaharada},
  {Kawano}, {Kawashima}, {Kawasoe}, {Kishishita}, {Kitaguchi}, {Kobayashi},
  {Kokubun}, {Kotoku}, {Kouda}, {Kubota}, {Kuroda}, {Madejski}, {Makishima},
  {Masukawa}, {Matsumoto}, {Mitani}, {Miyawaki}, {Mizuno}, {Mori}, {Mori},
  {Murashima}, {Murakami}, {Nakazawa}, {Niko}, {Nomachi}, {Okada}, {Ohno},
  {Oonuki}, {Ota}, {Ozawa}, {Sato}, {Shinoda}, {Sugiho}, {Suzuki}, {Taguchi},
  {Takahashi}, {Takahashi}, {Takeda}, {Tamura}, {Tamura}, {Tanaka}, {Tanihata},
  {Tashiro}, {Terada}, {Tominaga}, {Uchiyama}, {Watanabe}, {Yamaoka},
  {Yanagida}, \& {Yonetoku}}]{T07}
{Takahashi} T. {et~al.}, 2007, \pasj, 59, 35

\bibitem[{{Tanaka}(1986)}]{T86}
{Tanaka} Y., 1986, in Lecture Notes in Physics, Berlin Springer Verlag, Vol.
  255, IAU Colloq. 89: Radiation Hydrodynamics in Stars and Compact Objects,
  {Mihalas} D., {Winkler} K.-H.~A., eds., p. 198

\bibitem[{{Titarchuk}(1994)}]{TI94}
{Titarchuk} L., 1994, \apj, 434, 570

\bibitem[{{White} \& {Pravdo}(1979)}]{WP79}
{White} N.~E., {Pravdo} S.~H., 1979, \apjl, 233, L121

\bibitem[{{White}, {Swank} \& {Holt}(1983){White}, {Swank}, \& {Holt}}]{W83}
{White} N.~E., {Swank} J.~H., {Holt} S.~S., 1983, \apj, 270, 711

\end{thebibliography}
\bibliographystyle{mn2e}

\end{document}